\documentclass[
prc,showpacs,
preprintnumbers,amsmath,amssymb,floatfix,nofootinbib]{revtex4}

\usepackage{graphicx}
\usepackage{dcolumn}
\usepackage{bm}

\usepackage{amsmath}
\usepackage{psfrag}
\usepackage{color}
\usepackage{latexsym,amsfonts}
\usepackage{graphicx}
\providecommand{\href}[2]{#2}   

\def\txt#1{\textstyle{#1}}

\begin{document}

\title{Box diagram in the elastic electron-proton scattering}
\author{Dmitry~Borisyuk}
\author{Alexander~Kobushkin}%
\email{kobushkin@bitp.kiev.ua}
\affiliation{Bogolyubov Institute for Theoretical Physics\\
Metrologicheskaya str. 14-B, 03143, Kiev, Ukraine}

\date{\today}

\begin{abstract}
We present an 
evaluation of box diagram for the elastic $ep$ scattering with proton in the intermediate state. Using analytic properties of the proton form factors we express the amplitude via twofold integral, which involves the form factors in the space-like region only. Therefore experimentally measured form factors can be used in the calculations directly. The numerical calculation is done with the form factors extracted by Rosenbluth, as well as by polarization transfer methods. The dependence of the results on the form factor choice is small for  $Q^2 \lesssim 6 {\rm\ GeV}^2$, but becomes sizable at higher~$Q^2$. 
\end{abstract}

\maketitle
\section{Introduction}
Study of the elastic electron-proton scattering   
\begin{equation}
 e^- + p \to e^- + p
\end{equation}
is an important source of information about the proton structure. In the first order of the perturbation theory (PT) or Born approximation
the reaction amplitude
\begin{equation} \label{1ph}
 {\cal M}_\mathrm{Born} = {4\pi\alpha\over q^2} \bar u'\gamma_\mu u \cdot \bar U' \Gamma^\mu(q) U
\end{equation}
is expressed in terms of Dirac and Pauli proton form factors (FFs), $F_1$ and $F_2$,
\begin{equation} \label{Gamma}
 \Gamma_\mu(q) = F_1(q^2) \gamma_\mu -
  F_2(q^2) [\gamma_\mu,\gamma_\nu] {q^\nu \over 4M},
\end{equation}
where $u$, $U$, $u'$, $U'$ are 4-spinors of incoming and outgoing particles,
$\alpha \approx 1/137$ is fine structure constant, $q$ is the momentum transfer
from the electron to the proton.
The linear combinations, $G_E = F_1 + {q^2 \over 4M^2} F_2$ and
$G_M = F_1 + F_2$, called the electric and magnetic FFs, are also widely used.

\begin{figure}[t]
\centering
\psfrag{q1}{$q_1$}\psfrag{q2}{$q_2$}
\psfrag{p2}{proton}
\psfrag{delta}{$\Delta(1232)$}
\psfrag{=}{\Large =}
\psfrag{+}{\Large +}\psfrag{++}{\Large +\ \dots\ +}
\psfrag{Nonres}{\ non-resonant}
\psfrag{a}{(a)}\psfrag{b}{(b)}\psfrag{c}{(c)}\psfrag{d}{(d)}\psfrag{e}{(e)}
\includegraphics[width=0.75\textwidth]{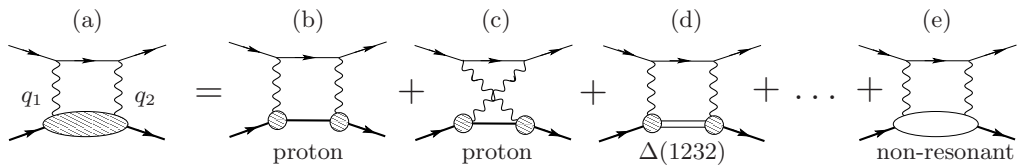}
\caption{Two photon exchange for the elastic $ep$ scattering. }
\label{sbox}
\end{figure}

During many years it was a common practice to extract FFs by the Rosenbluth separation method. FFs obtained by this method obey with good accuracy $G_E/G_M = {\rm const}$ for $0<Q^2 \lesssim 6 {\rm\ GeV}^2$ ($Q^2 \equiv -q^2$).
However since 2000 series of experiments was done using an alternative, polarization transfer method \cite{JlabExp}. These experiments yielded strikingly different results: $G_E/G_M$ ratio decreased linearly with $Q^2$. Since both methods are based on the Born approximation, the reason for discrepancy is likely the neglected higher order PT terms. 
There are two types of second order Feynman diagrams: first, diagrams, involving an  exchange of only one virtual photon (vacuum polarization or vertex corrections) and second, two photon exchange or box diagram, Fig.\ref{sbox}(a). The part of amplitude, coming from one-photon exchange diagrams, has the structure similar to (\ref{1ph}) (certainly with another functions in place of $F_{1,2}$) and cannot lead to the discrepancy between two methods. Such diagrams are usually taken into account in experimental analyses.

Therefore the two photon exchange diagram plays the key role in understanding the experimental results and extracting correct FF values.

The lower part of the diagram represents the doubly virtual Compton scattering
(VVCS) off the proton. The amplitude of VVCS has two poles which are due to single proton in the intermediate state. The contribution of these poles to the
box diagram is called the elastic contribution. Similar contributions
appear from $\Delta(1232)$ and other resonances, Fig.\ref{sbox}.
In the present paper we study the elastic contribution only (diagrams Fig.\ref{sbox}(b) and (c), which we later on call box (in the narrow sense) and x-box diagrams).
We also believe that the method derived here can be applied, with minor 
modifications, to the contributions of resonances.
This will be the subject of a separate paper.

In previously published papers the box diagram was calculated in several,
more or less approximate ways.

First is the soft photon approximation used in Tsai's paper \cite{MoTsai}
(and also in  Ref.\cite{MT}).
It is assumed that the main contribution to the loop integral comes from the
$q_1 \approx 0$, $q_2 \approx q$ and $q_1 \approx q$, $q_2 \approx 0$ regions.
The reason is usually given that the integrand
is strongly peaked at that points because of the infrared (IR) singularity.
However, only the IR divergent part (which exactly cancels in the final
answer), can be calculated in such a way. Instead, we are interested here
in the IR finite part, and it is by no means obvious, that the main
contribution to it comes from the same regions.

The second way, which may be called ``hard photon approximation'', assumes the most important
region to be $q_1 \approx q_2 \approx q/2$ \cite{Kuraev}. This may be a reasonable
assumption, however we believe it needs justification. In Ref.\cite{Kuraev},
after making the approximation the loop integral becomes divergent and an ultraviolet cut-off is inserted by hand.

In the third group of papers the loop integral was
calculated exactly, however using the special FF parameterization, monopole FF
in Ref.\cite{BMT0} and the sum of monopoles in Ref.\cite{BMT},
\begin{equation} \label{BMTfit}
 F_{1,2}(t) = \sum_i {n_i \over d_i - t},
\end{equation}
where $n_i$ and $d_i$ are fitted constants. Thanks to this simple form,
the resulting integrals are expressed through 4- and 3-point functions,
which were calculated numerically using computer program.

However, such sort of calculation has the following problem.
Since the FFs, entering the integral, are not known from the first
principles, we should use some model for them or fit to experiment
\footnote{Of course, this is a vicious circle, since the accurate FF extraction
from experiment implies taking into account the box diagram. However as a
zeroth approximation we may neglect it; then some iterative procedure may
be used to obtain precise result.}.
The problem is that only $t<0$ region is accessible for the
scattering experiments. While the $t>4M^2$ region can, in principle,
be studied in $e^- + e^+ \to p + \tilde p$ reaction, the unphysical region
$0<t<4M^2$ is completely inaccessible. But the loop integral involves
FFs in the time-like region ($t>0$) as well. A natural question arises, how the
(largely unknown) FF behaviour for $t>0$ influences the total
result.
In particular, parameterization used in Ref.\cite{BMT} does not even roughly reproduce FFs in the $t>0$ region.
Indeed, unitarity requires that FFs have poles at $t = m_V^2$ where $m_V$
are masses of vector mesons, namely $\rho$-, $\omega$- and $\varphi$- mesons
(neglecting their widths). But the numerical values given in Ref.\cite{BMT}
are away from these masses. So, the fit is not suitable for $t>0$ and the results of Ref.\cite{BMT} cannot be considered reliable before answering the above question.

In the present paper we calculate the box and x-box diagrams
in the most rigorous way. First, contrary to \cite{MoTsai,MT,Kuraev} we
evaluate all integrals exactly, without any restriction of the integration
domain. Second, contrary to \cite{BMT0,BMT} we do not use any special assumptions
on the FFs functional form. We perform the analytical integration
to the maximal possible extent, resulting in
\begin{equation} \label{res}
 {\cal M}_{\genfrac{}{}{0pt}{}{\rm box}{\rm x-box}} = \sum_{i,j=1}^2 \int_{t_1,t_2\le 0} {\cal K}_{ij}(t_1,t_2) F_i(t_1) F_j(t_2) dt_1 dt_2,
\end{equation}
where ${\cal K}_{ij}(t_1,t_2)$ is explicitly known.
During the derivation of (\ref{res}) we perform Wick rotation of the integration path, which is possible due to FF analyticity. As a result, the integration is done over negative $t_1$ and $t_2$ only.
Therefore we overcome the above-mentioned problem
of finding out FF values at $t>0$. Now the experimentally measured FFs can be
used directly for the calculation of (\ref{res}).
Then we calculate two photon exchange amplitudes numerically using different FF
parameterizations and discuss the results.
\begin{figure}[b]
\centering
\psfrag{k}{$k$}\psfrag{k1}{$k'$}\psfrag{k2}{$k''$}
\psfrag{q1}{$q_1$}\psfrag{q2}{$q_2$}
\psfrag{p}{$p$}
\psfrag{p1}{$p'$}
\psfrag{p2}{$p''$}
\includegraphics[width=0.25\textwidth]{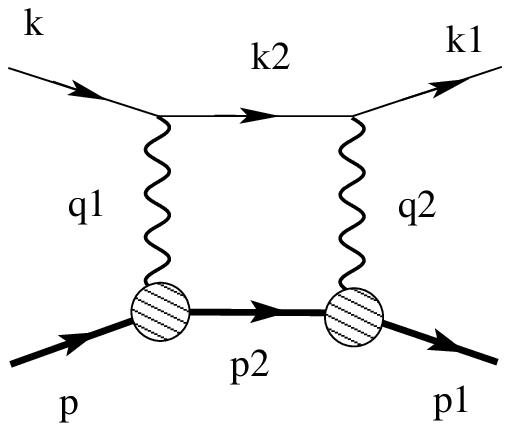}
\includegraphics[width=0.25\textwidth]{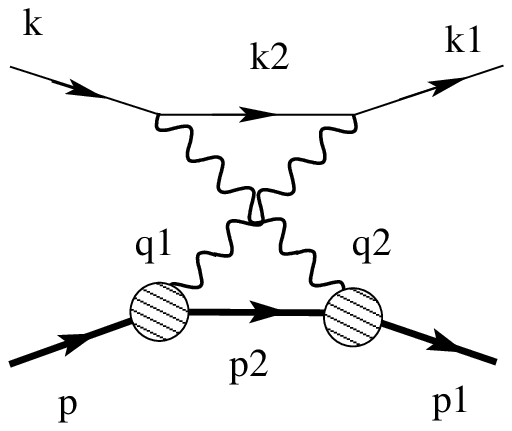}
\caption{Box diagram (left) and x-box diagram (right) with the proton in the intermediate state.}
\label{Twophot}
\end{figure}
\section{The method \label{sec2}}
%
%
The box and x-box diagrams, as well as notation for particle momenta, are displayed
in Fig.\ref{Twophot}. Thin line is for electron, thick is for proton.
The proton and electron masses are $M$ and $m$, respectively. We also denote
\begin{eqnarray}
 P = (p+p')/2, & K = (k+k')/2, & q = p'-p.
\end{eqnarray}
The following relations are fulfilled:
\begin{eqnarray}
 q_1 = p''-p, & q_2 = p''-p', & k'' = K \pm (P-p''),
\end{eqnarray}
upper sign is for the box, lower is for the x-box diagram.

The amplitude corresponding to either of diagrams is
\begin{equation} \label{intbox}
 i {\cal M}_{\genfrac{}{}{0pt}{}{\rm box}{\rm x-box}} = \left ({\alpha\over\pi}\right)^2 \int {N(p'') d^4 p'' \over
 (q_1^2-\lambda^2)(q_2^2-\lambda^2)(k''^2-m^2)(p''^2-M^2)},
\end{equation}
where
\begin{equation} \label{N}
 N(p'') = \bar u' \gamma_\mu (\hat k''+m) \gamma_\nu u \cdot
 \bar U' \Gamma_\mu(q_2) (\hat p''+M) \Gamma_\nu(q_1) U,
\end{equation}
$\Gamma_\mu$ is defined by (\ref{Gamma}) and we use the notation $\hat a \equiv a_\mu \gamma^\mu$.
For the x-box diagram one should interchange $\gamma_\mu$ with $\gamma_\nu$
in Eq.(\ref{N}). The ``photon mass'' $\lambda$
is introduced in the denominator to avoid IR divergence.
Though the electron mass is small compared to the characteristic energies
involved, we will not neglect it in the denominator, first, for generality,
second, since, as it is seen from the result, each of the diagrams
diverges like $\ln m$ as $m \to 0$ (but their sum does not).

In general case the elastic $ep$ scattering is described by six invariant amplitudes. However three of them are proportional to $m$, so in the $m\to 0$ limit remain only three \cite{Amp}:
\begin{equation}
 {\cal M} = {4\pi\alpha \over q^2} \bar u'\gamma_\mu u \cdot
 \bar U' \left(\tilde F_1 \gamma^\mu - \tilde F_2 
 [\gamma^\mu,\gamma^\nu] {q_\nu \over 4M} +
 \tilde F_3  \hat K {P^\mu \over M^2}
 \right) U.
\end{equation}
The invariant amplitudes $\tilde F_i$ depend on two kinematic variables,
$\nu = 4PK$ and $t=q^2$. In the first order of PT
$\tilde F_1(\nu,t) = F_1(t)$, $\tilde F_2(\nu,t) = F_2(t)$, $\tilde F_3(\nu,t) = 0$.
The contribution of the box diagram to any of these amplitudes is given
by the integral, similar to (\ref{intbox}), but with another numerator,
\begin{equation} \label{num}
 N(p'') = \sum_{i,j=1}^2 A_{ij}(p'') F_i(q_1^2) F_j(q_2^2),
\end{equation}
where $A_{ij}(p'')$ are some (rather cumbersome) explicitly known scalar
polynomial functions of $p''$. To simplify the notation, from now on we drop
the sum sign and the summation indices in the expressions like (\ref{num})
and write them simply as $A(p'') F(q_1^2) F(q_2^2)$. However the summation
is always understood. 

There are at most four independent scalar functions of $p''$, say, $p''^2$,
$pp''$, $p'p''$ and $Kp''$ (the pseudoscalar combination
$\epsilon_{\mu\nu\sigma\tau} p''^\mu p^\nu p'^\sigma K^\tau$ cannot appear
because of the parity considerations). The alternative, more convenient choice
is
\begin{equation} \label{vars}
\begin{array}{l}
  p''^2 - M^2,  \\
  k''^2 - m^2 =  \txt{t_1+t_2-t\over 2} \pm 2KP \mp 2Kp'',\\
  t_1 \equiv q_1^2 =  p''^2 + M^2 - 2pp'', \\
  t_2 \equiv q_2^2 =  p''^2 + M^2 - 2p'p'', 
\end{array}
\end{equation}
upper sign is for the box, lower is for the x-box diagram.
Since $A_{ij}(p'')$ are polynomials in $p''$, they can also be written as polynomials
in four variables (\ref{vars}).

Now it is easy to see that the integral (\ref{intbox}) can be reduced to the four
basic integrals:
\begin{equation} \label{int2}
 I_n = \int {A(t_1,t_2) \bar F(t_1) \bar F(t_2) \over D_n} d^4 p'',
\end{equation}
where $A$ is a polynomial, $\bar F(t) = F(t)/(t-\lambda^2)$, and
\begin{equation}
\begin{array}{rcl}
 D_1 & = & 1\\
 D_2 & = & k''^2-m^2\\
 D_3 & = & p''^2-M^2\\
 D_4 & = & (k''^2-m^2)(p''^2-M^2).
\end{array}
\end{equation}
%
%
%
If the maximal power of $k''^2-m^2$ or $p''^2-M^2$ in $A(p'')$ is greater
than one, then other integrals can appear in the decomposition of
(\ref{intbox}). However, they are expressed through (\ref{int2}) using
symmetry considerations, see Appendix~\ref{AppA}.

For the x-box diagram the integrals $I_1$, $I_2$, $I_3$, are the same as for
the box diagram, but the integral $I_4$ is not, since the relation between
$p''$ and $k''$ is different. We denote the corresponding integral $I_{\rm 4x}$.

We will show here in detail the evaluation of the integral $I_4$.
Other integrals are evaluated similarly. At the end we write down the results
for all of them.

As it was discussed previously, the idea is to integrate over two of four
integration variables, to obtain
\begin{equation} \label{result}
 I_n = \int {\cal K}_n(t_1,t_2) A(t_1,t_2) \bar F(t_1) \bar F(t_2) dt_1 dt_2
\end{equation}
with an explicitly known functions ${\cal K}_n$. This expression can be used for
numerical integration or further analysis.

\subsection*{Evaluation of ${\cal K}_n$}

We will perform the calculation in the Breit frame. In this frame we have
\begin{equation}
\begin{array}{ccc}
 P = ({1\over 2}\sqrt{4M^2-t},0,0,0) ,&
 q = (0,0,0,\sqrt{-t}),&
 K = ({\nu \over 2\sqrt{4M^2-t}}, \sqrt{\nu^2-(4m^2-t)(4M^2-t) \over 4(4M^2-t)},0,0)
\end{array} 
\end{equation}
(the first is the time component, the following are $x$-, $y$-, and $z$- components).
Let also
\begin{equation}
 p'' = (\xi + \txt{1\over 2}\sqrt{4M^2 - t}, \rho \cos \phi, \rho \sin \phi, \eta),
\end{equation}
so $d^4 p'' = d\xi d\eta\, \rho d\rho\, d \phi$. In this notation
\begin{eqnarray}
 t_{1,2} & = & \xi^2 - \rho^2 - ( \eta \pm \sqrt{-t}/2 )^2, \nonumber \\
 p''^2 - M^2 & = & \left(\xi + \txt{1\over 2}\sqrt{4M^2 - t} \right)^2 
 - \eta^2 - \rho^2 - M^2, \\
 k''^2 - m^2 & = & \left( \txt{\nu \over 2\sqrt{4M^2-t}} - \xi \right)^2
 - \rho^2 - \eta^2 - m^2 - K_x^2 + 2 \rho K_x \cos \phi. \nonumber 
\end{eqnarray}
We are to calculate the following integral:
\begin{equation}
 I_4 = \int { A(t_1,t_2) \bar F(t_1) \bar F(t_2) \over
 (p''^2-M^2)(k''^2-m^2) } \, d\xi \, d\eta\, \rho d\rho\, d\phi.
\end{equation}
First we integrate over $\phi$. The only quantity that does depend on
$\phi$ is $k''^2$. Using
\begin{equation} \label{defsqrt}
 {1 \over 2\pi} \int_0^{2\pi} {d\phi \over z-\cos \phi} =
 {1\over\sqrt{z^2-1}}
\end{equation}
it is easy to verify that
%
%
\begin{equation} \label{Kphi}
 D_\phi = \left({1 \over 2\pi} \int_0^{2\pi} {d\phi \over k''^2-m^2}\right)^{-1} = \sqrt{\left( \xi^2- \txt{\nu \over \sqrt{4M^2-t}}\xi -
 \rho^2 - \eta^2 - {t \over 4} \right)^2 - 4 K_x^2 \rho^2} .
\end{equation}
%
In Eq.(\ref{defsqrt}) and further the following convention is used:
we mean by $\sqrt{z^2-a^2}$ the analytic function of $z$ with the branch cut
from $-a$ to $a$ such that $\sqrt{z^2-a^2} > 0$ for $z>a$ and
$\sqrt{z^2-a^2} < 0$ for $z<-a$. Consequently, Eq.(\ref{Kphi}) implies, that
if $D_\phi$ is real, it has the same sign as the expression in the brackets under the radical.
The integral becomes
\begin{equation} \label{defPhi}
I_4 = 2\pi \int { A(t_1,t_2) \bar F(t_1) \bar F(t_2) \over
 (p''^2-M^2) D_\phi(\xi,\eta,\rho) } \, d\xi \, d\eta \, \rho d\rho =
 \int \Phi \, d\xi \, d\eta \, d\rho^2.
\end{equation}

\begin{figure}[t]
\centering
\psfrag{1}{\it 1}\psfrag{2}{\it 2}\psfrag{3}{\it 3}\psfrag{4}{\it 4}\psfrag{k}{$\xi$}
\includegraphics[width=0.6\textwidth]{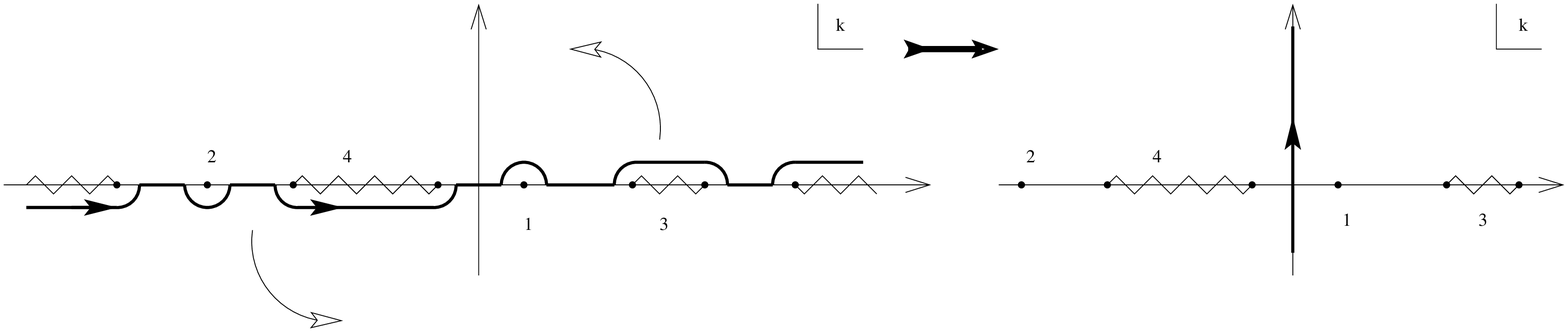}
\psfrag{1}{\it 1}\psfrag{2}{\it 2}\psfrag{3}{\it 3}\psfrag{4}{\it 4}
\includegraphics[width=0.6\textwidth]{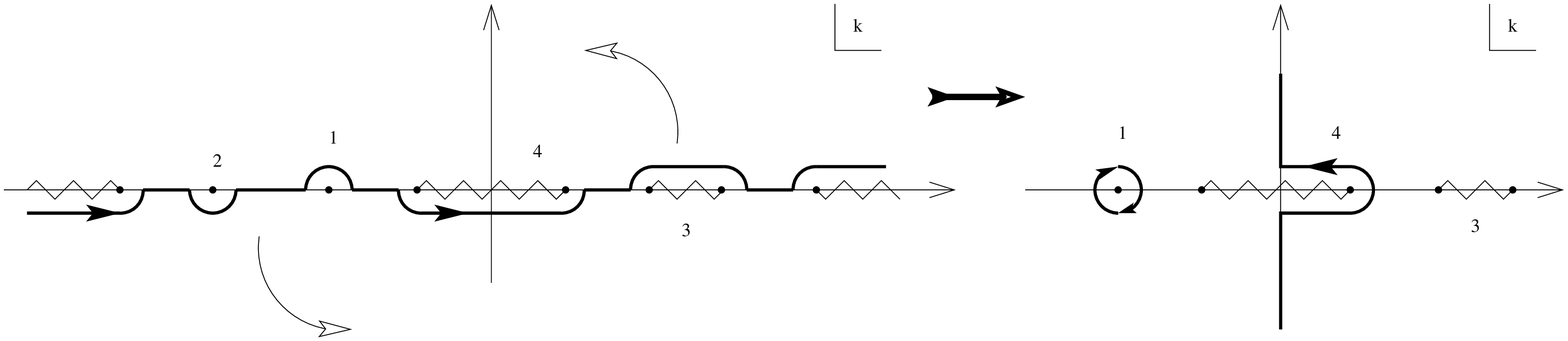}
\caption{The integration path before (left) and after Wick rotation (right).}
\label{path}
\end{figure}

Next we will integrate over $\xi$. Here we intend to use Wick rotation, so
we need to study the analytic properties of the integrand. The FFs
$F(t)$ (and thus $\bar F(t)$) are analytic functions of $t$ everywhere in the physical sheet except
the positive real axis. Namely, FFs have branch cuts running from $t=(2m_\pi)^2$
to $+\infty$ ($m_\pi$ is the pion mass) \cite{FFprop}, and the additional pole at $t=\lambda^2$
is due to the photon propagator; the physical value is $\bar F(t+i0)$.
In the complex $\xi$ plane the corresponding singularities lie at $\xi^2 > \rho^2+(|\eta|-\sqrt{-t}/2)^2$.
The factor $1/(p''^2-M^2)$ has two poles {\it 1} and {\it 2} at
$ \xi = -{1\over 2}\sqrt{4M^2-t} \pm  \sqrt{\rho^2 + \eta^2 + M^2} $ and
$D_\phi$ has two branch cuts {\it 3} and {\it 4} running from
$ \xi = {\nu \over 2\sqrt{4M^2-t}} \pm \sqrt{(\rho-K_x)^2+\eta^2+m^2} $
to
$ {\nu \over 2\sqrt{4M^2-t}} \pm \sqrt{(\rho+K_x)^2+\eta^2+m^2} $.
When integrating over $\xi$, we must pass by these singularities
adding $-i0$ to the masses in the usual way. The resulting integration path
$\ell$ is shown in Fig.\ref{path} on the left.

Now we perform the Wick rotation to superpose the integration path with
the imaginary axis. If the singularities {\it 1} and {\it 3} lie at $\xi>0$ and
{\it 2} and {\it 4} lie at $\xi<0$, then the integration path can be rotated without
crossing them, Fig.\ref{path}, upper drawing. While singularities {\it 2} and {\it 3} always fulfill the
above condition, it may not hold for {\it 1} and {\it 4}. In this case an extra
terms appear when the path crosses the singularities, Fig.\ref{path}, lower drawing.

In general case we may write
\begin{eqnarray} \label{int1}
 \int_\ell \Phi d\xi = \int_{-i\infty}^{+i\infty} \Phi \, d\xi +
 \int \Delta\Phi \, d\xi \, \theta(\xi)
 \theta \left( \txt{\nu \over 2\sqrt{4M^2-t}} - \xi \right) \theta(-D_\phi^2) - \\
 - 2\pi i \int [(p''^2-M^2)\Phi] \, \theta(-\xi)
 \theta \left( \xi+\txt{{1 \over 2}\sqrt{4M^2-t}} \right) \delta(p''^2-M^2) \, d\xi, \nonumber
\end{eqnarray}
where $\Delta\Phi = \Phi(\xi-i0)-\Phi(\xi+i0) = 2\Phi(\xi-i0)$.
It can be verified that during the integration according to (\ref{int1}) $t_1$
and $t_2$ are always negative. For the first integral in the r.h.s.
$\xi^2<0$ and thus $t_{1,2} = \xi^2 - \rho^2 - (\eta \pm \sqrt{-t}/2)^2 < 0 $.
As we will see later, the same is true for other integrals.
This is a great advantage, since the FFs are well-known for $t<0$
(which corresponds to the elastic scattering), and much worse known
for $t>0$.

Here we will change variables to make new independent variables
($t_1$, $t_2$, $\xi$) instead of ($\rho$, $\eta$, $\xi$). It is important
to take into account that $\rho^2 > 0$; for this purpose we insert
$\theta(\rho^2)$ under the integral. This implies the condition:
$\xi^2 > -{t\over 4}+{t_1+t_2\over 2}-{(t_1-t_2)^2 \over 4t} = \xi_*^2$, and the
integration path along the imaginary axis becomes bounded.

It is convenient to introduce the notation
\begin{equation}
\begin{array}{l}
x_\infty = {1\over t} \left({t_1 + t_2 -t \over 2}\right)^2 - 
  {1 \over t}\, t_1 t_2, \\
x_M = {1\over t} \left({t_1 + t_2 -t \over 2}\right)^2 - 
  \left({1 \over t} - {1 \over 4M^2}\right) t_1 t_2, \\
x_m = {1\over t} \left({t_1 + t_2 -t \over 2}\right)^2 -
   \left({1 \over t} - {1 \over 4m^2}\right) t_1 t_2.  
\end{array}
\end{equation}
Note that at $t_1,t_2 < 0$ one has $x_m > x_M > x_\infty$.

In new variables we have
\begin{eqnarray}
 p''^2-M^2 & = & \xi \sqrt{4M^2-t} + \txt{t_1+t_2-t\over 2} =
  \sqrt{4M^2-t} (\xi - \xi_0 - C), \\
 \label{Kphinew} 
 D_\phi & = & -\sqrt{4m^2-t}\sqrt{(\xi-\xi_0)^2-B^2},
\end{eqnarray}
where $\xi_0 = {t_1+t_2-t\over2}{\nu\over(4m^2-t)\sqrt{4M^2-t}}$,
$B = {4 m K_x \over 4m^2-t} \sqrt{x_m}$,
$C = -{t_1+t_2-t \over 2\sqrt{4M^2-t}} \left( 1+{\nu \over 4m^2-t} \right)$.
Therefore when $\Phi$ is expressed through $t_1$, $t_2$ and $\xi$, it has, as a function of $\xi$, just one pole and one branch cut. 
It is worth noting that (\ref{Kphinew}) may differ in sign from
$D_\phi$ defined according to (\ref{Kphi}). However along the integration
path they are always equal.
\begin{figure}[t]
\centering
\psfrag{k}{$\xi$}\psfrag{ks}{$\xi_\ast$}\psfrag{ksm}{$-\xi_\ast$}
\psfrag{b1}{}\psfrag{b2}{}\psfrag{c}{}
\includegraphics[width=0.25\textwidth]{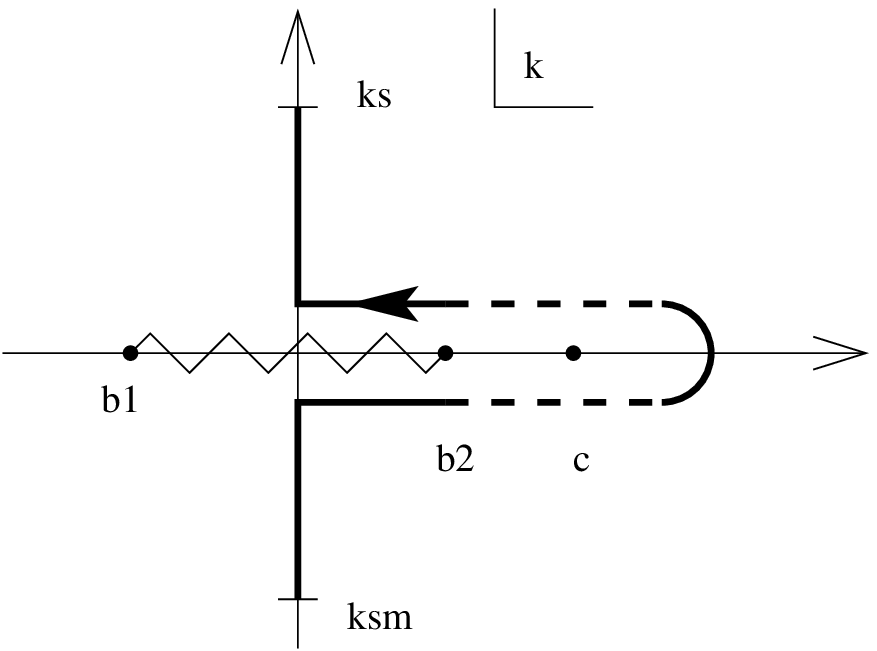}
\caption{New integration path $\ell'$.}
\label{path3}
\end{figure}

Rewriting (\ref{int1}) in new variables and after long algebraic
transformations (see Appendix~\ref{AppB}),
we obtain
\begin{eqnarray} \label{inttt}
 I_4 = -\pi \!\!\!\!\! \intop_{t_1,t_2 \le 0} \!\! {dt_1 dt_2 \over 2\sqrt{-t}}
 {A(t_1,t_2) \bar F(t_1) \bar F(t_2) \over \sqrt{4m^2-t} \sqrt{4M^2-t}}
 \left\{ \theta(x_\infty)
  \int_{\ell'} {d\xi \over (\xi-\xi_0-C)\sqrt{(\xi-\xi_0)^2-B^2}} + \right. \\
  + 2 i \, \theta(x_m) \theta(-x_\infty)\theta(t_1+t_2-t)
   \int_{\xi_0-B}^{\xi_0+B} {d\xi \over (\xi-\xi_0-C)\sqrt{B^2-(\xi-\xi_0)^2}}
  - \nonumber \\ \left. - 2\pi i
   [ \theta(x_\infty) + \theta(x_M)\theta(-x_\infty)\theta(t_1+t_2-t) ]
   {1\over \sqrt{C^2-B^2}}
 \right\}, \nonumber
\end{eqnarray}
where $\ell'$ is the integration path depicted in Fig.\ref{path3}.

The integration over $\xi$ now can be done analytically.
The second integral, after substitution $\xi = \xi_0 + B \cos\varphi$,
is done with the help of Eq.(\ref{defsqrt}) and gives
\begin{equation}
 \int_{\xi_0-B}^{\xi_0+B} {d\xi \over (\xi-\xi_0-C)\sqrt{B^2-(\xi-\xi_0)^2}} =
 \int_0^\pi {d\varphi \over B\cos \varphi - C} = -{\pi \over \sqrt{C^2-B^2}}.
\end{equation}
The first integral is done using
\begin{equation} \label{log}
 \int {dz \over (z-C)\sqrt{z^2-B^2}} = {1\over 2 \sqrt{C^2-B^2}} \ln
 {Cz-B^2-\sqrt{C^2-B^2}\sqrt{z^2-B^2} \over Cz-B^2+\sqrt{C^2-B^2}\sqrt{z^2-B^2}}
 = \Lambda(z).
\end{equation}
Here and below $\ln z $ is defined with the branch cut from $z=0$ to $+\infty$,
so that for $z>0$ ${\rm Im } \ln(z+i0) = 0$ and ${\rm Im } \ln(z-i0) = 2\pi$.

It is easy to show that the argument of logarithm in Eq.(\ref{log}) can be 
real positive only if $z$ is real, so all the discontinuities of $\Lambda(z)$
lie on the real axis. Therefore
\begin{eqnarray}
 \int_{\ell'} {d\xi \over (\xi-\xi_0-C)\sqrt{(\xi-\xi_0)^2-B^2}} 
 = \Lambda(\xi_*-\xi_0) - \Lambda(-\xi_*-\xi_0) -
 \Lambda(+\infty+i0) + \Lambda(+\infty-i0). \nonumber 
\end{eqnarray}
Using
\begin{eqnarray}
 & \sqrt{(\xi_*-\xi_0)^2-B^2} = \frac{1}{\sqrt{4m^2-t}} \left( {\nu\xi_* \over \sqrt{4M^2-t}} - {t_1+t_2-t \over 2} \right) &
\end{eqnarray}
%
%
%
after some simplifications we arrive at the final result of the form
(\ref{result}) with ${\cal K}_4$ which is given below together with other
${\cal K}_n$.
\begin{eqnarray}
 {\cal K}_1(t_1,t_2) & = & {\pi \xi_* \over \sqrt{-t}} \theta(x_\infty),  \\
 {\cal K}_2(t_1,t_2) & = &
  {\pi \over 2\sqrt{-t(4m^2-t)}} \left\{ \theta(x_\infty) 
   \left. \ln \left( \xi + \txt{t_1+t_2-t \over 2\sqrt{4m^2-t}}
     \right) \right|_{-\xi_*}^{\xi_*} -
     2\pi i \, \theta(x_m)\theta(-x_\infty)\theta(t_1+t_2-t) \right\}, \\
 {\cal K}_3(t_1,t_2) & = & 
  {\pi \over 2\sqrt{-t(4M^2-t)}} \left\{ \theta(x_\infty) 
   \left. \ln \left( \xi + \txt{t_1+t_2-t \over 2\sqrt{4M^2-t}}
    \right) \right|_{-\xi_*}^{\xi_*} -
    2\pi i \, \theta(x_M)\theta(-x_\infty)\theta(t_1+t_2-t) \right\}, \\
 {\cal K}_4(t_1,t_2) & = & 
  {\pi \over 2\sqrt{ R_+ }} \left\{ \theta(x_\infty)
   \left. \ln \left[ \tfrac{\xi}{\sqrt{-t}} \sqrt{ R_+ } - \nu x_\infty
   + \txt{\left(t_1+t_2-t \over 2\right)^2} \right]\right|_{-\xi_*}^{\xi_*}
   + \right. \nonumber \\
   & + & \left. 2\pi i \, \theta(x_\infty) + 2\pi i \,
   \theta(-x_\infty) \theta(t_1+t_2-t) [\theta(x_m)+\theta(x_M)] \right\}, \\
 {\cal K}_{\rm 4x}(t_1,t_2) & = & 
  {-\pi \over 2 \sqrt{ R_- }} \left\{ \theta(x_\infty)
   \left. \ln \left[ \tfrac{\xi}{\sqrt{-t}} \sqrt{ R_- } - \nu x_\infty
   - \txt{\left(t_1+t_2-t \over 2\right)^2} \right]\right|_{-\xi_*}^{\xi_*}
   + \right. \nonumber \\
  & + & \left. 2\pi i \, \theta(t_1+t_2-t) [\theta(x_m)\theta(-x_M) +
   2\theta(R_-)\theta(x_M)\theta(\nu+t-4M^2)] \right\},   
\end{eqnarray}
where $\xi_* = i\sqrt{x_\infty}$ and
\begin{equation}
 R_\pm = \left( \txt{t_1+t_2-t \over 2}\right)^2
 ((\nu \mp t)^2 - 16 m^2 M^2 ) -
  t_1 t_2 (\nu^2- (4m^2-t)(4M^2-t)),
\end{equation}
$\sqrt{R_\pm} = \sigma |R_\pm|^{1/2}$, where $\sigma = - {\rm\, sign\, } (t_1+t_2-t)$ if $R_\pm>0$ and $\sigma = -i$ if $R_\pm<0$.
\section{Numerical}
It is convenient to distinguish three integration regions: i) $x_\infty \ge 0$
ii) $x_M \ge 0 \ge x_\infty$ and iii) $x_m \ge 0 \ge x_M$.
In general case the integrals $I_2$, $I_4$ and $I_{\rm 4x}$ contain logarithmic terms
like $\ln {m \over \lambda}$ originating from the integration over
$x_m \ge 0 \ge x_M$ region. However in actual calculations after adding box and x-box
diagrams these logarithms always cancel;
it can be shown to be the consequence of gauge invariance.
Therefore we may put $m=0$ if the amplitude is evaluated as a whole (box + x-box).

Before calculating $ep \to ep$ amplitudes, the following cross-check can be performed.
If we omit $A(t_1,t_2)$ and set $F(t) \equiv 1$ in Eq.(\ref{int2}) then
the integrals $I_n$ can also be done by usual Feynman parameters method.
We have calculated them numerically with different values of 
$M$, $m$, $\lambda$, $t$, $\nu$, and  made sure that both methods give
identical results.

Now we turn to evaluation of invariant amplitudes $\tilde F_1$, $\tilde F_2$,
$\tilde F_3$. Each of them looks like
\begin{equation}
 \tilde F_i = a_i \ln \lambda + b_i + o(\lambda).
\end{equation}
The first (IR divergent) term does not contribute to the cross-section if
the radiation of soft photons is taken into account.
To extract both IR divergent and IR finite part and to simplify
the calculation we used the following procedure. We calculated the integrals
needed at several small but non-zero $\lambda$ and fitted obtained values
with 3-parameter fit:
\begin{equation}
 \tilde F_i = a_i \ln \lambda + b_i + c_i \lambda.
\end{equation}
Though the third term vanishes as $\lambda \to 0$, it turns out to be
necessary to obtain accurate results.

The results obtained by Tsai \cite{MoTsai} are
\begin{equation} \label{resMT}
 \tilde F_{1,2}^{(T)} = {\alpha \over \pi} F_{1,2}
 \left[ K(p,k') - K(p,k) \right] ,\ \ \ \tilde F_3^{(T)} = 0,
\end{equation}
where $K(p_i,p_j) = (p_i \cdot p_j) \int_0^1 {dy \over p_y^2} \ln {p_y^2\over \lambda^2}$,
$p_y = p_i y + p_j (1-y)$; the superscript (T) hereafter indicates Tsai's result.
In spite of the approximate nature of this result, the terms proportional to
$\ln\lambda$ are exact; they are
\begin{equation}
 a_{1,2} = a_{1,2}^{(T)} = {2\alpha \over \pi} F_{1,2} \ln {\nu-t \over \nu+t}.
\end{equation}
Our numerical calculation has given the same results for these terms.

Instead of $\tilde F_{1,2}$ we introduce linear combinations $\tilde G_M$ and
$\tilde G_E$ exactly as this is done for FFs, $\tilde G_E =
\tilde F_1 + {t\over 4M^2} \tilde F_2$ and $\tilde G_M = \tilde F_1 + \tilde F_2$.
To see the relative size of corrections with respect to the Born approximation,
we consider the following quantities:
\begin{eqnarray}
 \delta G_M / G_M & = & (\tilde G_M - G_M^{(T)} - G_M ) / G_M, \nonumber \\
 \delta G_E / G_M & = & (\tilde G_E - G_E^{(T)} - G_E ) / G_M, \\
 Y_{2\gamma} & = & (\nu / 4M^2) ( \tilde F_3 / G_M ). \nonumber 
\end{eqnarray}
The results of Tsai are subtracted, since they were used
(the therefore already taken into account) in the analysis of almost all experimental data.
This is not the same as just drop the logarithmic term, since
(\ref{resMT}) contains non-logarithmic terms as well.
\begin{figure}[t]
\centering
\includegraphics[width=0.47\textwidth]{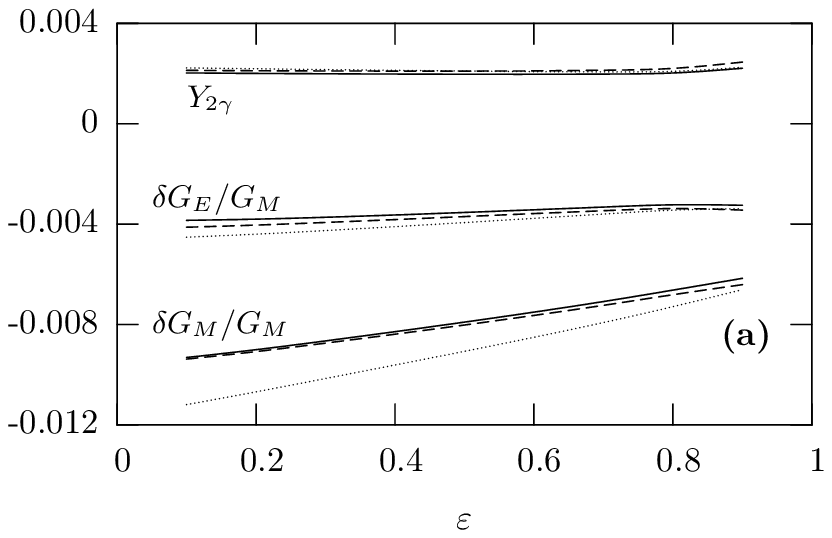}
\includegraphics[width=0.47\textwidth]{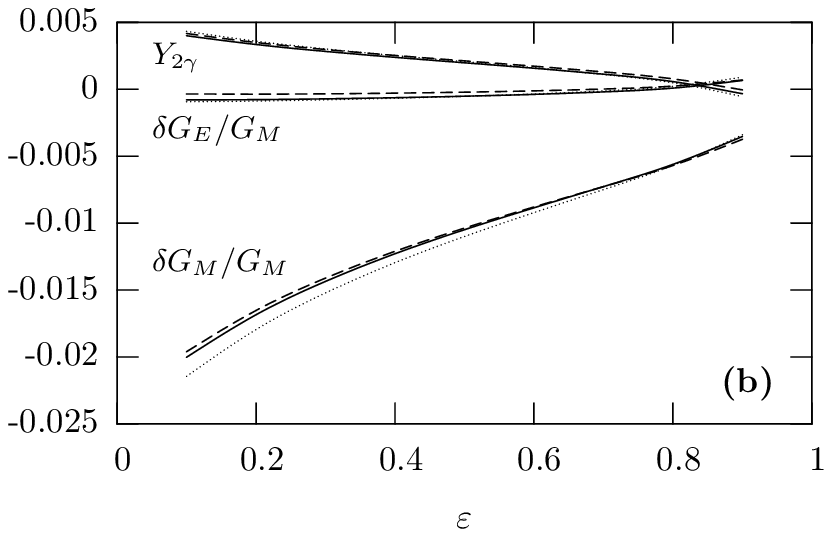}
\includegraphics[width=0.47\textwidth]{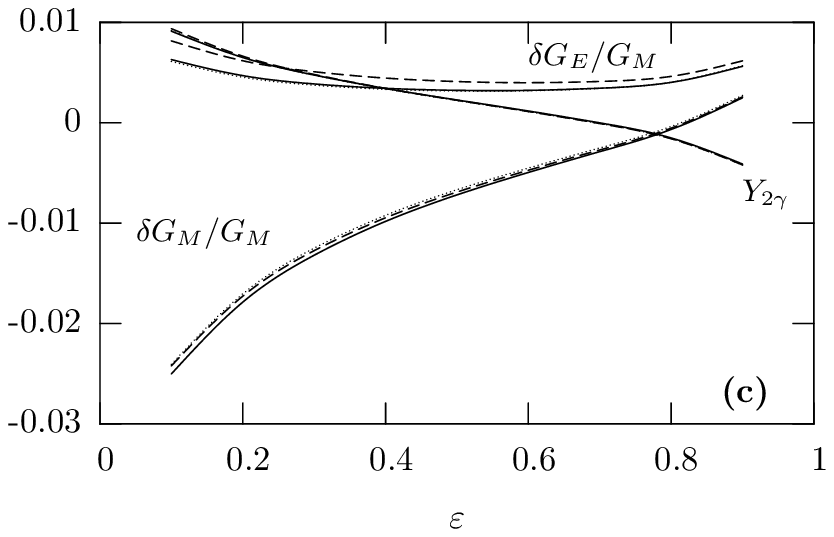}
\includegraphics[width=0.47\textwidth]{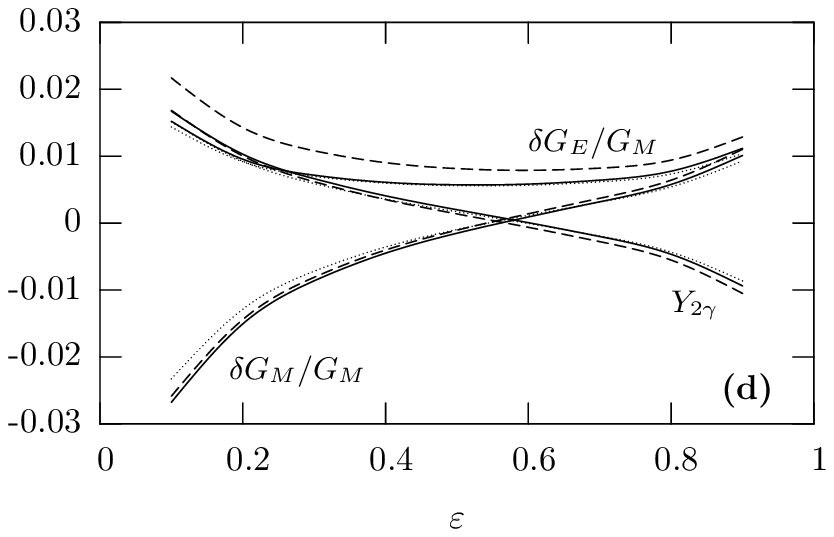}
\caption{Two photon exchange corrections $\delta G_M/G_M$, $\delta G_E/G_M$ and $Y_{2\gamma}$ as indicated on the figures, for $Q^2$ (a) 1 GeV$^2$, (b) 3 GeV$^2$, (c) 6 GeV$^2$, (d) 10 GeV$^2$, using form factor parameterization: from Ref.\protect\cite{Bosted} (solid), from Ref.\protect\cite{JlabFit} (dashed) and dipole fit (dotted).}
\label{curves}
\end{figure}

Note that in contrast to Ref.\cite{BMT}, we normalize $\delta G_E$
by $G_M$, rather than $G_E$. This is done because of great discrepancy
in $G_E$ as extracted by two methods; in addition, $G_E$ as extracted by
polarization transfer method is close to zero at $Q^2 \approx 6{\rm\ GeV}^2$
and $\delta G_E/G_E$ becomes large though $\delta G_E$ itself is small.
On the other hand, the uncertainty in $G_M$ is much smaller.

We have calculated the two photon exchange amplitudes at $Q^2=1$,3,6
and 10 $\rm GeV^2$ using three FF parameterizations:
\begin{itemize}
\item{dipole fit}
\item{fit from Ref.\cite{Bosted} (under assumption $G_E/G_M = \rm const$)}
\item{fit from Ref.\cite{JlabFit} (to data obtained by polarization transfer)}
\end{itemize}
The results are plotted in Fig.\ref{curves} versus 
customary parameter $\varepsilon$ (virtual photon polarization), $\varepsilon = {\nu^2 + t(4M^2-t) \over \nu^2 - t(4M^2-t)}$.

We see that in general results agree with those obtained in Ref.\cite{BMT}.
This is not surprising, because in Sec.\ref{sec2} we actually prove that given the FF parameterization which is good for space-like region ($t<0$) and has correct analytic properties, the result will not depend on the
quality of the fit for $t>0$ ({\it a priori} this was not clear).
In some sense we justify the approach of Ref.\cite{BMT}. Nevertheless, our approach is more general, since we can use any FF parameterization.


We also see that the dependence on the choice of FFs is small for $Q^2<6\rm\ GeV^2$, however for the generalized electric FF at $Q^2 = 10 \rm\ GeV^2$ the difference reaches 50\% (for small $\varepsilon$).

The absolute value of corrections increases with $Q^2$ up to $\sim 3\%$ for
$\delta G_M/G_M$ at $Q^2=10 \rm\ GeV^2$.

The main purpose of this paper is to present new approach for evaluation
of two photon exchange diagram. That is why we will not analyze the effect of
these corrections on the FF measurements. Such analysis requires evaluation of
bremsstrahlung corrections as well, and thus detailed consideration
of experimental conditions is needed.
We postpone it to another paper.
 

\section{Summary}

We study the two photon exchange in the elastic $ep$ scattering. In this paper we 
calculate box and crossed-box diagrams with proton in the intermediate state (elastic contribution).
Our approach has two main advantages:
\begin{itemize}
 \item the amplitude is expressed via proton form factors in the space-like region only, Eq.(\ref{res}). The previous calculations required form factors in the time-like region as well, where they are poorly known.
 \item any suitable form factor parameterization in the space-like region can be used to evaluate the loop integral. In previous calculations only particular kind of parameterization could be used, since the loop integral was expressed via 4-point functions.
\end{itemize}

Similar approach can be used to evaluate $\Delta(1232)$ and other resonances contributions,
which is currently under investigation. At present only the $\Delta$ contribution
was considered in the literature \cite{Delta} using approach of Refs.\cite{BMT0,BMT}
and simple dipole parameterization of the $N\to \Delta$ transition form factors.

We have calculated two photon exchange amplitudes using form factors as extracted by Rosenbluth and by polarization transfer methods. The dependence of the result on the choice of form factors is small except the generalized electric form factor for $Q^2 \ge 6 \rm\ GeV^2$.

The same method can be applied to the scattering off any spin-1/2 particle, such as neutron or $^3$He.

\appendix
\section{}\label{AppA}
In this Appendix we show how some integrals, that can appear in the decomposition of (\ref{intbox}), are reduced to four basic types (\ref{int2}).

Consider the integral
\begin{eqnarray}
& \int A(t_1,t_2) {P''^2-M^2 \over k''^2-m^2} d^4p'' = 
\int A(t_1,t_2) {(P+K-k'')^2 - M^2 \over k''^2-m^2} d^4p'' = & \\
& = {\nu + 4m^2 - t \over 2} \int A(t_1,t_2) {1 \over k''^2-m^2} d^4p'' + 
 \int A(t_1,t_2) d^4p'' - & \nonumber \\
& - 2 (P+K)^\mu \int A(t_1,t_2) {k''_\mu \over k''^2-m^2} d^4p''. & \nonumber
\end{eqnarray}
The first and the second integral in the r.h.s. already have required form.
Since $t_{1,2} = (K \pm q/2 - k'')^2$, the last integral depends only on $q^\mu$ and $K^\mu$, thus
\begin{eqnarray}
 & \int A(t_1,t_2) {k''_\mu \over k''^2-m^2} d^4p'' = 
 \alpha q_\mu + \beta K_\mu, &
\end{eqnarray}
where
\begin{eqnarray}
 \txt{\alpha = {1 \over 2 t} \int A(t_1,t_2) {t_2-t_1 \over k''^2-m^2} d^4p''} & &
 \txt{\beta = {1 \over 4m^2 - t} \int A(t_1,t_2)
   \left({4m^2 - t_1-t_2 \over k''^2-m^2} + 2 \right) d^4p'',}
\end{eqnarray}
so finally
\begin{eqnarray}
 & \int A(t_1,t_2) {p''^2-M^2 \over k''^2-m^2} d^4 p'' = 
  {1\over 2} \left(1+{\nu \over 4m^2-t} \right)
  \int A(t_1,t_2) {t_1+t_2-t \over k''^2-m^2} d^4 p'' -
  {\nu \over 4m^2-t} \int A(t_1,t_2) d^4 p''. &
\end{eqnarray}
Similarly one obtains
\begin{eqnarray}
 & \int A(t_1,t_2) {k''^2-m^2 \over p''^2-M^2} d^4 p'' = 
  {1\over 2} \left(1+{\nu \over 4M^2-t} \right)
  \int A(t_1,t_2) {t_1+t_2-t \over p''^2-M^2} d^4 p'' -
  {\nu \over 4M^2-t} \int A(t_1,t_2) d^4 p'' &
\end{eqnarray}
and 
\begin{eqnarray}
 & \int A(t_1,t_2) (k''^2-m^2) d^4 p'' = 
   \int A(t_1,t_2) (p''^2-M^2) d^4 p'' = 
   \int A(t_1,t_2) {t_1+t_2-t \over 2} d^4 p''. &
\end{eqnarray}

\section{}\label{AppB}
In this Appendix we show in detail the derivation of Eq.(\ref{inttt}).

Let us transform two last integrals in Eq.(\ref{int1}).
It follows from Eq.(\ref{Kphi}) that if $D_\phi^2<0$ then always $\rho^2>0$,
so $\theta(\rho^2)\theta(-D_\phi^2) = \theta(-D_\phi^2)$, and also
\begin{equation}
 {\left. D_\phi^2 \right|}_{\xi = {\nu \over 2\sqrt{4M^2-t}}} = 
 (\rho^2+m^2+\eta^2-K_x^2)^2 + 4K_x^2(m^2+\eta^2) > 0,
\end{equation}
so $\theta \left( {\nu \over 2\sqrt{4M^2-t}} - \xi \right)$ is constant 
(either 1 or 0) when $D_\phi^2 < 0$. Therefore
\begin{eqnarray}
 & & \theta(\rho^2) \theta(\xi) 
  \theta \left( \txt{\nu \over 2\sqrt{4M^2-t}} - \xi \right)
  \theta(-D_\phi^2) = \nonumber \\
 & & = \theta(\xi)\theta \left( \txt{\nu \over 2\sqrt{4M^2-t}} - \xi_0 \right)
  \theta( B^2-(\xi-\xi_0)^2) = \\
 & & = \theta(\xi)\theta(4m^2-t_1-t_2)\theta(x_m)\theta( B^2-(\xi-\xi_0)^2). \nonumber
\end{eqnarray}
$\theta(x_m)$ can be inserted since $0 < B^2-(\xi-\xi_0)^2$ implies
$0 < B^2 = \left( {4mK_x \over 4m^2-t} \right)^2 x_m$.
\begin{figure}[t]
\centering
\psfrag{t1}{$t_1$}\psfrag{t2}{$t_2$}\psfrag{t}{$t$}
\psfrag{4m}{$4m^2$}
\includegraphics[width=0.3\textwidth]{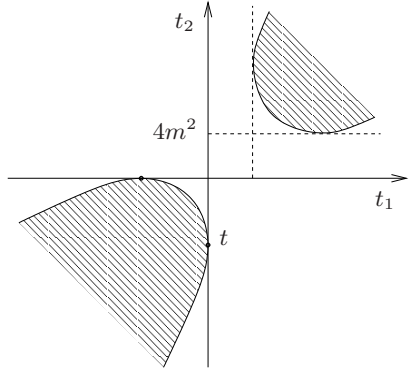}
\caption{Regions in $(t_1,t_2)$ plane, where $x_m>0$ (hatched).}
\label{hyperbola}
\end{figure}
Rewriting $x_m$ as
\begin{eqnarray}
 & {4m^2 \over 4m^2-t} x_m = 
 {1\over 4m^2-t} \left( t_1+t_2-4m^2 \over 2 \right)^2 +
 {1 \over t} \left( t_1-t_2 \over 2 \right)^2 - m^2 &
\end{eqnarray}
it is easy to see that $x_m$ is positive in two regions, bounded by hyperbola
$x_m = 0$ (Fig.\ref{hyperbola}). The condition $4m^2-t_1-t_2>0$ selects lower-left region,
where $t_1,t_2 \le 0 $. Thus
\begin{equation} \label{t<0}
 \theta(4m^2-t_1-t_2)\theta(x_m) = \theta(-t_1)\theta(-t_2)\theta(x_m)
\end{equation}
and
\begin{equation}
 \theta(\rho^2) \theta(\xi)
  \theta \left( \txt{\nu \over 2\sqrt{4M^2-t}} - \xi \right) \theta(-D_\phi^2) =
 \theta(-t_1)\theta(-t_2)\theta(x_m)\theta(\xi)\theta(B^2-(\xi-\xi_0)^2).
\end{equation}
To transform the last integral, we write
\begin{eqnarray}
 & & \theta(\rho^2) \theta(-\xi)
  \theta \left( \xi+\txt{1\over 2}\sqrt{4M^2-t} \right) \delta(p''^2-M^2) = \\
 & & = \theta(\xi^2+x_\infty) \theta(-\xi) 
  \theta \left( \xi+\txt{1\over 2}\sqrt{4M^2-t} \right)
  \delta \left( \xi\sqrt{4M^2-t} + \txt{t_1+t_2-t \over 2} \right) = \nonumber \\
 & & = \theta(x_M) \theta(t_1+t_2-t) \theta(4M^2-t_1-t_2) \txt{1\over\sqrt{4M^2-t}}
  \delta \left( \xi + \txt{t_1+t_2-t \over 2\sqrt{4M^2-t}} \right) = \nonumber \\
 & & = \theta(-t_1) \theta(-t_2) \theta(x_M) \theta(t_1+t_2-t)
  \txt{1\over\sqrt{4M^2-t}}
  \delta \left( \xi + \txt{t_1+t_2-t \over 2\sqrt{4M^2-t}} \right). \nonumber  
\end{eqnarray}
The last equality was obtained using Eq.(\ref{t<0}) with $m$ replaced by $M$.
Again we have $t_1,t_2 \le 0$.

After that the integral $I_4$ becomes
\begin{eqnarray} \label{prev}
 I_4 = \intop_{t_1,t_2 \le 0} {dt_1 dt_2 \over 2\sqrt{-t}} \left\{
 \int_{-i\infty}^{+i\infty} \theta(\xi^2+x_\infty) \, \Phi \, d\xi +
 \theta(x_m) \int \Delta\Phi \, d\xi \, \theta(\xi) \theta(B^2-(\xi-\xi_0)^2) - \right. \\
 \left. - {2\pi i \over \sqrt{4M^2-t}} \theta(t_1+t_2-t) \theta(x_M)
 \int [(p''^2-M^2)\Phi] \, \delta
 \left( \xi+ \txt{t_1+t_2-t \over 2\sqrt{4M^2-t}}\right) d\xi \right\}. \nonumber
\end{eqnarray}
Now we introduce the integration path $\ell'$ that passes by the
singularities on the right, Fig.\ref{path3}. After that
\begin{eqnarray}
 \int_{-i\infty}^{+i\infty} \theta(\xi^2+x_\infty) \, \Phi \, d\xi =
 \theta(x_\infty) \left\{ \int_{\ell'} \Phi \, d\xi - 
 \int \Delta\Phi \, d\xi \, \theta(\xi) \theta(B^2-(\xi-\xi_0)^2) - \right. \\
 \left. - 2\pi i \, \theta(\xi_0+C)
 \int [(\xi - \xi_0 - C)\Phi] \, \delta (\xi-\xi_0-C) \, d\xi \right\}. \nonumber
\end{eqnarray}
Combining this with the previous equation, we have
\begin{eqnarray}
 I_4 = \intop_{t_1,t_2 \le 0} {dt_1 dt_2 \over 2\sqrt{-t}}
 \left\{ \theta(x_\infty) \int_{\ell'} \Phi \, d\xi + 
  \theta(x_m) \theta(-x_\infty)
   \int \Delta\Phi \, \theta(\xi) \, \theta(B^2-(\xi-\xi_0)^2) d\xi - \right. \\
  \left. - 2\pi i
   [ \theta(x_\infty)\theta(t-t_1-t_2) + \theta(x_M)\theta(t_1+t_2-t) ]
   \int [(\xi - \xi_0 - C) \Phi] \,
   \delta (\xi - \xi_0 - C) \, d\xi
 \right\}. \nonumber
\end{eqnarray}
To proceed further, we note that
\begin{equation}
 {\left. B^2-(\xi-\xi_0)^2 \right|}_{\xi = 0} =
 \txt{\nu^2 \over (4m^2-t)(4M^2-t)}x_\infty - \txt{4m^2\over 4m^2-t} x_m < 0
\end{equation}
for $x_\infty < 0$ and $x_m > 0$. Therefore
\begin{eqnarray}
 & & \theta(x_m)\theta(-x_\infty)\theta(\xi)\theta(B^2-(\xi-\xi_0)^2) = \\
 & & = \theta(x_m)\theta(-x_\infty)\theta(\xi_0)\theta(B^2-(\xi-\xi_0)^2) = \nonumber \\
 & & = \theta(x_m)\theta(-x_\infty)\theta(t_1+t_2-t)\theta(B^2-(\xi-\xi_0)^2) \nonumber.
\end{eqnarray}
Also
\begin{eqnarray}
 & & \theta(x_\infty)\theta(t-t_1-t_2) + \theta(x_M)\theta(t_1+t_2-t) = \\
 & & = \theta(x_\infty) + (\theta(x_M)-\theta(x_\infty))\theta(t_1+t_2-t) = \nonumber\\
 & & = \theta(x_\infty) + \theta(x_M)\theta(-x_\infty) \theta(t_1+t_2-t) \nonumber
\end{eqnarray}
and finally
\begin{eqnarray} \label{int3}
 I_4 = \intop_{t_1,t_2 \le 0} {dt_1 dt_2 \over 2\sqrt{-t}}
 \left\{ \theta(x_\infty) \int_{\ell'} \Phi \, d\xi + 
  \theta(x_m) \theta(-x_\infty)\theta(t_1+t_2-t)
   \int \Delta\Phi \, \theta(B^2-(\xi-\xi_0)^2) d\xi - \right. \\
  \left. - 2\pi i
   [ \theta(x_\infty) + \theta(x_M)\theta(-x_\infty)\theta(t_1+t_2-t) ]
   \int [(\xi-\xi_0-C) \Phi] \,
   \delta ( \xi-\xi_0-C ) \, d\xi
 \right\}. \nonumber
\end{eqnarray}
Substituting, according to (\ref{defPhi})
\begin{eqnarray}
 \Phi = {- \pi A(t_1,t_2) \bar F(t_1) \bar F(t_2) \over \sqrt{4m^2-t} \sqrt{4M^2-t}}
  \ { 1 \over (\xi - \xi_0 - C) \sqrt{ (\xi-\xi_0)^2 - B^2 }}, \\
 \Delta\Phi = {- \pi A(t_1,t_2) \bar F(t_1) \bar F(t_2) \over \sqrt{4m^2-t} \sqrt{4M^2-t}}
  \ { 2i \over (\xi - \xi_0 - C) \sqrt{ B^2 - (\xi-\xi_0)^2}},
\end{eqnarray}
into (\ref{int3}), we obtain Eq.(\ref{inttt}).

\end{document}